\documentclass[%
 pre,
rsi,%
 amsmath,amssymb,
 reprint,%
superscriptaddress
]{revtex4-1}

\usepackage{graphicx}
\usepackage{dcolumn}
\usepackage{bm}
\usepackage{color}

\begin {document}

\def\bea{\begin{eqnarray}}
\def\eea{\end{eqnarray}}
\def\beq{\begin{equation}}
\def\eeq{\end{equation}}
\def\f{\frac}
\def\k{\kappa}
\def\e{\epsilon}

\def\D{\Delta}
\def\h{\theta}
\def\t{\tau}
\def\a{\alpha}

\def\r{\rho}

\def\s{\sigma}
\def\kb{k_B}

\def\la{\langle}
\def\ra{\rangle}

\def\d{\delta}

\def\a{\alpha}
\def\d{\delta}

\def\la{\langle}
\def\ra{\rangle}
\def\e{\epsilon}

\title{Re-entrant phase separation in nematically aligning active polar particles}

\author{Biplab Bhattacherjee} 
\email{biplab@iopb.res.in}
\affiliation{Institute of Physics, Sachivalaya Marg, Bhubaneswar 751005, India.}
\author{Debasish Chaudhuri}
\email{debc@iopb.res.in}
\affiliation{Institute of Physics, Sachivalaya Marg, Bhubaneswar 751005, India.}
\affiliation{Homi Bhaba National Institute, Anushaktigar, Mumbai 400094, India.}

\date{\today}

\begin{abstract}
We present a numerical study of the phase behavior of repulsively interacting active polar particles that align their active velocities nematically. The amplitude of the active velocity, and the noise in its orientational alignment control the active nature of the system. At high values of orientational noise, the structural fluid undergoes a continuous nematic-isotropic transition in active orientation.
This transition is well separated from an active phase separation, characterized by the formation of high density hexatic clusters, observed at lower noise strengths.
With increasing activity, the system undergoes a re-entrant fluid- phase separation- fluid transition. The phase coexistence at low activity can be understood in terms of motility induced phase separation.  In contrast, the re-melting of hexatic clusters, and the collective motion at low orientational noise are dominated by flocking behavior. At high activity,  sliding and jamming of polar sub-clusters,  formation of grain boundaries, lane formation, and subsequent fragmentation of the polar patches mediate remelting.   
\end{abstract}
\maketitle

\section{Introduction}
Active particles transduce stored or ambient energy into motion,  generating non-equilibrium drive from the smallest scales of a system~\cite{Marchetti2013, Romanczuk2012, Vicsek2012}. Examples in biology range from molecular motors, cytoskeleton, cells, tissues~\cite{Schliwa2003, Alberts2007, Howarda, Surrey2001, Schaller2010}, bacterial suspension~\cite{Cates2012, Sokolov2007} to ant trails, bird flocks, animal groups and human crowd~\cite{Toner2005, Ballerini2008, Couzin2007, Couzin2003, Katz2011, Attanasi2014, Helbing1997, Helbing2000}. 
In nonliving matter, examples include colloidal Janus particles~\cite{Zottl2016}, and vibrated asymmetric granular matter~\cite{Weber2013}. 
The seminal model by Vicsek, and its subsequent extensions~\cite{Vicsek1995, Chate2006, Chate2008, Vicsek2012}, considered active point particles that orient the direction of their active velocity ferromagnetically, i.e.,  parallel to that of their neighbours. 
The collective properties of active particles depend only on a few microscopic details. 
The self propulsion can be polar~\cite{Vicsek1995, Gregoire2004} or apolar~\cite{Chate2006,Gruler1999}.
The direction of propulsion may align parallel to or in a nematic manner with the neighboring particles~\cite{Vicsek1995, Ginelli2010}. The alignment  may arise from  hydrodynamic coupling~\cite{Sokolov2007, Katz2011}, visual cues~\cite{Nagy2010, Ballerini2008}, or direct collision between extended  objects~\cite{Narayan2007, Peruani2006, Weber2013, Weitz2015, Shi2018, Grossman2019}. 
With increasing tendency of alignment and activity, the coupling of velocity fluctuations with density lead to flocking, collective motion, and giant number fluctuations~\cite{Toner1995, Toner1998, Narayan2007, Dey2012, Gregoire2004, Peruani2011b, Solon2013, Peruani2013}.

Another important aspect in the emergent phase behavior of active particles is the impact of volume exclusion.
Its interplay with activity is most extensively studied in active Brownian particles (ABP)~\cite{Fily2012, Redner2013,  Solon2015, Stenhammar2013, Cates2015}. In ABPs  the orientation of self propulsion of each particle undergoes Brownian rotation, independent of others. 
This persistent motion  suppresses the change in velocity direction after collision between ABPs, unlike in elastic scattering. The resultant slow-down generates a positive feedback effect allowing accumulation of more particles to form clusters, eventually leading to a  motility induce phase separation (MIPS)~\cite{Cates2015}.
 The high density clusters could be in either a hexatic or solid phase, determined by the amount of activity and mean density~\cite{Digregorio2018}. It has been shown recently that at low activity, the equilibrium two stage defect mediated  melting scenario of two-dimensional solid~\cite{Kosterlitz1973,  Halperin1978, Young1979, Kapfer2015, Bernard2011, Gasser2010, Thorneywork2017} persists, with  the melting points shifting to higher densities with increasing activity~\cite{Klamser2018, Digregorio2018}.

The two aspects of active  alignment and exclusion have been combined in
a number of recent studies analyzing the impact of volume exclusion on ferromagnetically aligning active polar particles~\cite{Peruani2011b, Weber2014, Martin-Gomez2018, Sese-Sansa2018}. The role of topological defects in the melting of such active solids was  demonstrated~\cite{Weber2014}. At lower densities, the system shows formation of micro or macro clusters, lanes and bands~\cite{Martin-Gomez2018}. 
A recent experiment observed interrupted phase separation in active Janus colloids where impact of alignment and excluded volume are simultaneously present~\cite{Linden2019}. 
 Active granular disks with built-in asymmetry showed polar activity under external shaking. 
 These disks repel each other in a circularly symmetric fashion, and their active velocities showed approximate nematic alignment~\cite{Weber2013}. 
 Active point particles with nematic alignment were known to undergo nematic to isotropic phase transition and associated clustering~\cite{Ginelli2010}.

In this paper, we investigate the effect of volume exclusion on the collective properties of nematically aligning active polar particles.
Using numerical simulations, we present a comprehensive study of the active orientational phase  transition and spatial phase separation with changing values of the active velocity and orientational noise. We fix the mean density at a value that is slightly lesser than the equilibrium melting point.
With reduction of alignment noise, the active fluid shows a continuous isotropic to nematic phase transition. However, across  this transition, structurally, the system remains in a homogeneous fluid phase, unlike the point particles.
With further reduction of the noise, the active particles  undergo a phase separation showing coexistence of high density hexatic clusters and low density fluid. 
This emerges due to the interplay of activity and excluded volume interaction. 
At an intermediate noise,  the system demonstrates a remarkable re-entrant transition from fluid - phase coexistence - fluid with increasing active velocity. 
We present the detailed phase diagram, supported by a comprehensive characterization of the respective phases and phase-transitions in this paper. 
As we show, the onset of clustering at low active velocities is like MIPS, mediated by an effective slowdown of particles with density. The remelting of these active hexatic clusters at higher active velocities takes place by formation of grain boundaries and fragmentation of local polar patches.  
The cluster size distribution changes qualitatively at the onset of the phase separation. 
The hexatic clusters display complex dynamics -- from sliding stripes, to jamming, to formation of counter-propagating channels. A finite size scaling analysis shows the robustness of the clustering transition, the mean cluster size remains a finite fraction of the overall system size even in the largest systems.

In Section~\ref{sec:model}, we describe the model and simulation details. 
We present the complete phase diagram and the details  of the phase transitions in Section~\ref{sec:results}.
Finally we conclude in Section~\ref{sec:discussion}  presenting a summary and outlook.

\section{Model}
\label{sec:model}
We consider a two-dimensional system of  $N$ active polar particles in a volume $A=L_x L_y$, performing active Brownian  motion with a self-propulsion speed $v_0$, in direction ${\textbf n_i} \,=(\cos\h_i, \sin\h_i)$ for the  $i$-th particle, where $\h_i$ denotes the angle ${\textbf n_i}$ subtends with the  $x$-axis. 
The positional evolution of the $i$-th particle is described by an overdamped Langevin equation
\begin{equation}
\dot{\bf r}_{i} = v_0 {\textbf n_i}(t) + 
\mu \sum_j {\bf F}_{ij} + 
\xi_i(t),
\label{eq02}
\end{equation}
where, $\mu$ is the mobility, $\xi_i(t)$ is a Gaussian white noise with $\langle \xi_i (t) \rangle = 0$ and $\langle \xi_i(t) \xi_i(0) \rangle = 2 D \delta(t)$ so that $D=\mu \kb T$,
${\bf F}_{ij} = -\nabla_i U(r_{ij})$, is the repulsive force between particles given by the WCA potential~\cite{Weeks1971}
\[
    U(r_{ij})=  
\begin{cases}
    4\e \left\lbrace  (\sigma/r_{ij})^{12}-(\sigma/r_{ij})^{6} \right\rbrace, & \text{if } r_{ij} \leq 2^{1/6}\s\\
    0,              & \text{otherwise}
\end{cases}
\]
where, $r_{ij} = \mid {\bf r}_i - {\bf r}_j \mid$ is the inter-particle separation. 
We assume a nematic alignment of the polar  orientation ${\textbf n_i}$  of the $i$-th particle with its neighbors. It aligns (anti-)~parallel to the local mean active orientation, if the relative angle between the two orientations is (obtuse) acute. 
We use the evolution~\cite{Ginelli2010},
\begin{equation}
\theta_i^{t+\delta t} = arg\left[ \sum_{j \in R_i} {\rm sign}[\cos(\theta_j^t - \theta_i^t)] e^{i\theta_j^t} \right] + \zeta_i(\eta).
\label{eq01}
\end{equation}
The alignment gets randomized by the orientational noise $\zeta_i(\eta)$ that follows uniform distribution between $(-\eta \pi, \eta \pi)$ with the noise strength $\eta \in (0,1)$. 

The unit of length, energy, and time are set by $\sigma$, $\e=k_B T$, and $\tau = \s^2/D$,  respectively.
The activity of the particles is controlled by a dimensionless P{\'e}clet number $Pe = v_0 \s/D$, and the noise strength $\eta$. The numerical simulation is performed using Euler integration. We use integration time step $\delta t=10^{-5}\tau$ for $Pe\leq 50$, and a smaller time-step $\delta t=5\times 10^{-6}\tau$ for higher $Pe$, to avoid  inter-particle overlap.

In the absence of activity, the two-dimensional  solid is expected to display a solid- hexatic- liquid equilibrium phase-transition~\cite{Bernard2011,Kapfer2015, Kosterlitz1973,Halperin1978,Young1979}, with the solid melting point at the dimensionless density $\r = (N/A) (\pi/4) \s^2 \approx 0.71$.
 We present results for $\r=0.6$, a density slightly below the equilibrium melting point. At this density, the excluded volume interaction plays an important role, while not dominating over activity.

\begin{figure*}[!t]
\begin{center}
\includegraphics[width=0.9\linewidth]{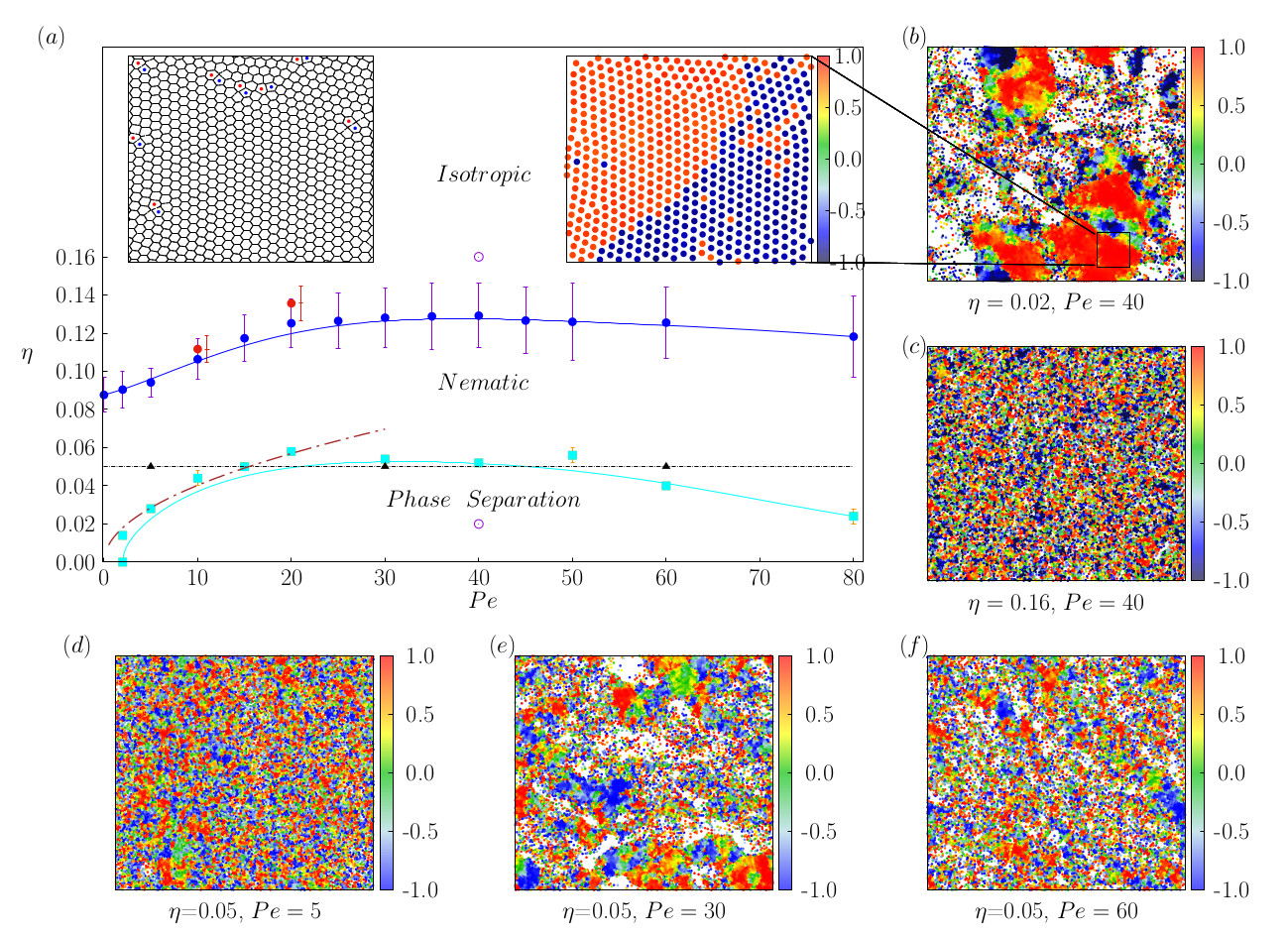} 
\end{center}
\caption{(color online)
($a$)~$Pe-\eta$ phase diagram at  $\r =0.6$:  The  filled blue circles show the isotropic-nematic phase boundary for a system of $N=1024$ particles. For comparison, we show two  isotropic-nematic transition points~(filled red circles) evaluated for a larger system of $N=16384$ particles.
The boundary of phase separation is shown by the cyan filled squares evaluated using the  $N=16384$ system. 
The solid lines through the two phase boundaries are guides to the eye.  The brown dash-dotted line shows the expected MIPS boundary, $\eta_c \sim Pe^{1/2}$ at smaller $Pe$. 
The black dotted line at a constant $\eta=0.05$ indicates  the re-entrant transition from a structural fluid ($Pe\lesssim 10$), to active phase coexistence, to again a single fluid phase ($Pe\gtrsim  60$). 
In ($b$)--($f$), some typical instantaneous density maps of  the local hexatic orientations are plotted with respect to the global hexatic orientation, $h_k = \vec \psi^k_6 \cdot \vec \psi_6$ for the $k$-th particle in the $N=16384$ system, at various $(Pe,\, \eta)$ points of the phase diagram. 
They show ($b$)~large hexatically correlated clusters deep inside  the phase coexistence region at $(\eta,Pe)=(0.02,40)$, and ($c$)~loss of that correlation in the isotropic phase at $(\eta,Pe)=(0.16,40)$. 
Plots ($d$)-($f$) show the same local hexatic orientations along the re-entrant transition line at $\eta=0.05$, and $Pe=5,\, 30,\, 60$, respectively (points denoted by $\blacktriangle$ on the re-entrant line in ($a$)\,). 
{ Insets in ($a$):} The right hand side inset shows particle positions from a region of high hexatic order in ($b$), with the color code here denoting the  orientations of active velocities of the  particles. 
The plot shows an approximate  triangular lattice like arrangement. 
The left hand side inset shows Voronoi tesselation corresponding to the same configuration identifying the bound $7-5$ disclination pairs denoted by the red and blue dots  respectively.
}
\label{PD}
\end{figure*}

\section{Phase transition}
\label{sec:results}
In this section, we present a detailed phase diagram, and extensive discussion of the corresponding phase transitions.
\subsection{Phase diagram}
\label{sec:ph_dia}
Fig.\ref{PD}($a$) shows the phase diagram in the $Pe$-$\eta$ plane, identifying the isotropic, and the nematic fluid, and the region of active phase separation. With decreasing $\eta$, the active orientations of isotropic fluid gets aligned in the nematic fashion, captured by the time-averaged scalar order parameter $S=\f{1}{\t_0}  \int_0^{\t_0} dt\,|{\bf s(t)}|$, where the mean nematic order of a given configuration is ${\bf s} = (1/N) \sum_{k=1}^N \exp(2i\h_k)$.  The  increasing convection of orientational information with increasing $Pe$, requires a larger $\eta$ to destroy the nematic order. However, due to the  volume exclusion, the range over which the nematic alignment may get extended due to convection is restricted at a given density. Thus the isotropic- nematic phase boundary shows an  initial increase with $Pe$ before eventual  saturation~(Fig.\ref{PD}($a$)\,).
The transition points denoted by the  filled blue circles in Fig.\ref{PD}($a$) are evaluated using $N=1024$ particles. They are identified by the maximum of fluctuations in the order parameter $S$.  For comparison, we performed simulations for a larger system of $N=16384$ particles at  $Pe=10$ and $20$. The red filled circles show the isotropic-nematic transition points corresponding to this system size. They compare well with $N=1024$ particle results, within error bars. The determination of the transition points and the corresponding error bars are discussed in detail in Sec.~\ref{sec:iso_nema}.

At further lower $\eta~(<0.06)$, the active system undergoes phase separation between a nematic fluid and  denser hexatic clusters. The phase coexistence is captured by the bimodal distribution function of the local density $\r({\bf r})$, and the  amplitude of the local hexatic bond-orientational order $\vec{\psi_l} ({\bf r}) =  \la \vec \psi_6^k\ra_{k \in {n_l}}$, where $n_l$ denotes the number of particles in a coarse-grained volume $l_x \times l_y$ around ${\bf r}$. The bond-orientational order associated with the $k$-th particle is defined as $\vec \psi_6^k = \sum_{j=1}^{n_v} (l_j/l) \, \exp(6i\phi_{kj})$~\cite{Mickel2013, Steinhardt1983}, where, $\phi_{kj}(t)$ is the angle subtended by the bond between the $k$-th particle and its $j$-th Voronoi neighbor on the positive $x$-axis;  
$l_j$ is the length of the Voronoi edge corresponding to the $j$-th 
neighbor, and $l = \sum_{j=1}^{n_v} l_j$ where $n_v$ is the total number of Voronoi topological neighbors. 
%
The solid line denotes the onset of phase coexistence of the nematic fluid and hexatic clusters. 
The non-monotonic variation of this boundary as a function of $Pe$ indicates a re-entrant transition. 
For example, keeping the orientational noise strength fixed at $\eta=0.05$, change in $Pe$ from $5$ to $30$ leads a nematic fluid to the hexatic- nematic fluid phase coexistence. 
The hexatic clusters remelt into a homogeneous nematic fluid by increasing $Pe$ further to $60$.
The dashed line  $\eta \sim Pe^{1/2}$ shown at the phase boundary is an estimate obtained using the argument of MIPS, as  outlined later in Sec.\ref{sec:mips}.
The clusters in the rest of the phase coexistence regime are dominated by alignment and flocking.  
Before going into the quantitative analysis, we illustrate the local phase behavior in terms of system   configurations and local hexatic order.


Fig.\ref{PD}($b$)-($f$) show typical instantaneous configurations of particles shaded by colors denoting the projection $h_k=\vec{\psi}_6^k\cdot \vec{\psi}_6$, of their individual hexatic orientations $\vec{\psi}_6^k$ to the instantaneous system averaged bond-orientational order $\vec{\psi}_6$. $h_k$ is bounded between $-1$ and $1$.
Fig.\ref{PD}($b$) shows a typical configuration in the two phase regime. The high density clusters of particles are observable as contiguous colored  patches. 
The large red patches identify regions with local bond orientational order aligned along  the mean instantaneous hexatic order. 
On the other hand, the blue patches denote the regions where local bond orientational orders are anti-parallel to that of the mean hexatic order. A magnified portion of one such high density cluster is shown in the upper right inset in Fig.\ref{PD}($a$). 
It shows local formation of solid-like triangular lattice structure. This region of high six-fold bond orientational order is a jammed state of two sets of active particles with opposite polarity, denoted by red and blue colors in this figure. 
In the upper left inset of Fig.\ref{PD}($a$), we show the same configuration in terms of Voronoi tessellation, color coding the disclinations where the particles do not have six  Voronoi neighbours, denoting particles with five neighbors in blue and that with seven neighbors in red. In a pure triangular lattice, all particles have six neighbors.  The presence of bound $5-7$ disclination pairs, well-separated from each other, identify free dislocations, a  characteristic of the hexatic phase.   
Fig.\ref{PD}($c$) shows a typical  configuration in the isotropic fluid phase. The two phase points at which configurations Fig.\ref{PD}($b$),($c$) are plotted are denoted by pink circles in the phase diagram  Fig.\ref{PD}($a$).
The black dashed line in the phase diagram Fig.\ref{PD}($a$) follows a remelting transition. Typical configurations corresponding to the three points denoted by $\blacktriangle$ in Fig.\ref{PD}($a$) are presented in Fig.\ref{PD}($d$)-($f$), showing significant hexatic clustering in Fig.\ref{PD}($e$).
The details of these phase transitions are quantified in the following.  

\begin{figure}[t]
\includegraphics[width=0.95\linewidth]{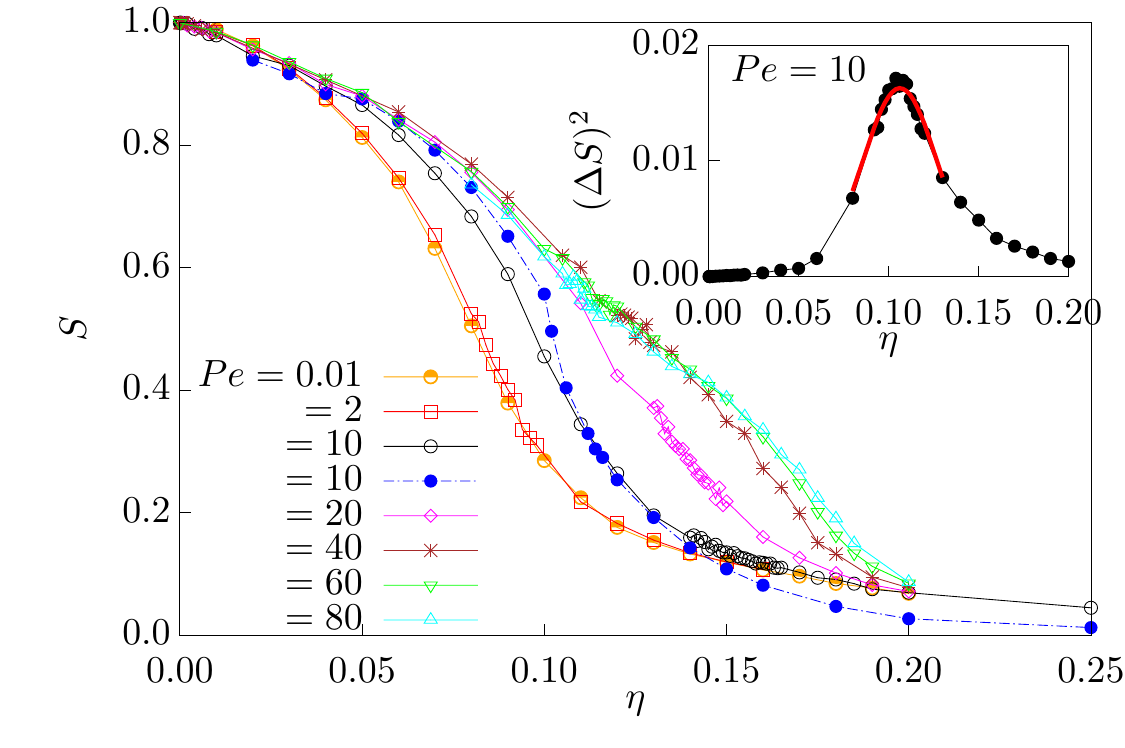}
\caption{(color online) Variations  of the mean nematic order parameter $S$ against the noise strength $\eta$ are shown at seven different P\'{e}clet numbers. The mean density of the system is fixed at $\r=0.6$. The variance in $S$, ${(\Delta S)}^2$, is shown at  $Pe=10$ in the inset. The central part of this curve is fitted with a Gaussian function (red solid line).  The position of the maximum of this Gaussian fit is identified as the transition point, and the standard deviation as the error bar. 
}
\label{NIPT1}
\end{figure}

\subsection{Isotropic Nematic Transition}
\label{sec:iso_nema}
The nematic-isotropic transition is quantified by following the reduction of the nematic scalar order parameter $S$ with the increase in the orientational noise strength $\eta$. This is shown for seven different $Pe$ values in Fig.\ref{NIPT1}, using a system of $N=1024$ particles. 
The fluctuation of the order parameter $(\Delta S)^2=[\la S^2\ra - \la S\ra ^2]$ shows a pronounced maximum at the transition point $\eta_c$. 
A representative behavior is shown in the inset of Fig.\ref{NIPT1} for $Pe=10$. 
This maximum is identified as the transition point. To determine the exact location of the maximum $\eta_c$ and its standard deviation $\D \eta_c$, we use a Gaussian fitting around the maximum~(red solid line in the inset of Fig.\ref{NIPT1}). 
These constitute the nematic-isotropic phase  transition points and their error-bars plotted  in Fig.\ref{PD}($a$).  
With increasing system size, the transition becomes sharper, however, the transition point remains within the error bar. One data set for a larger system of $N=16384$ particles at $Pe=10$ is shown  using the filled blue circles in Fig.\ref{NIPT1}. 
The observed transition is between a quasi-long ranged ordered (QLRO) nematic phase to a disordered isotropic  phase. The QLRO  within the nematic phase is illustrated in Appendix-\ref{sec_app:B} in terms of an algebraic decay of the scalar order parameter with system size.
A further discussion on the nature of $S(\eta)$ curves, and their data-collapse in an intermediate $Pe$ regime are presented in Appendix-\ref{sec_app:A}.  

\begin{figure}[t]
\includegraphics[width=0.95\linewidth]{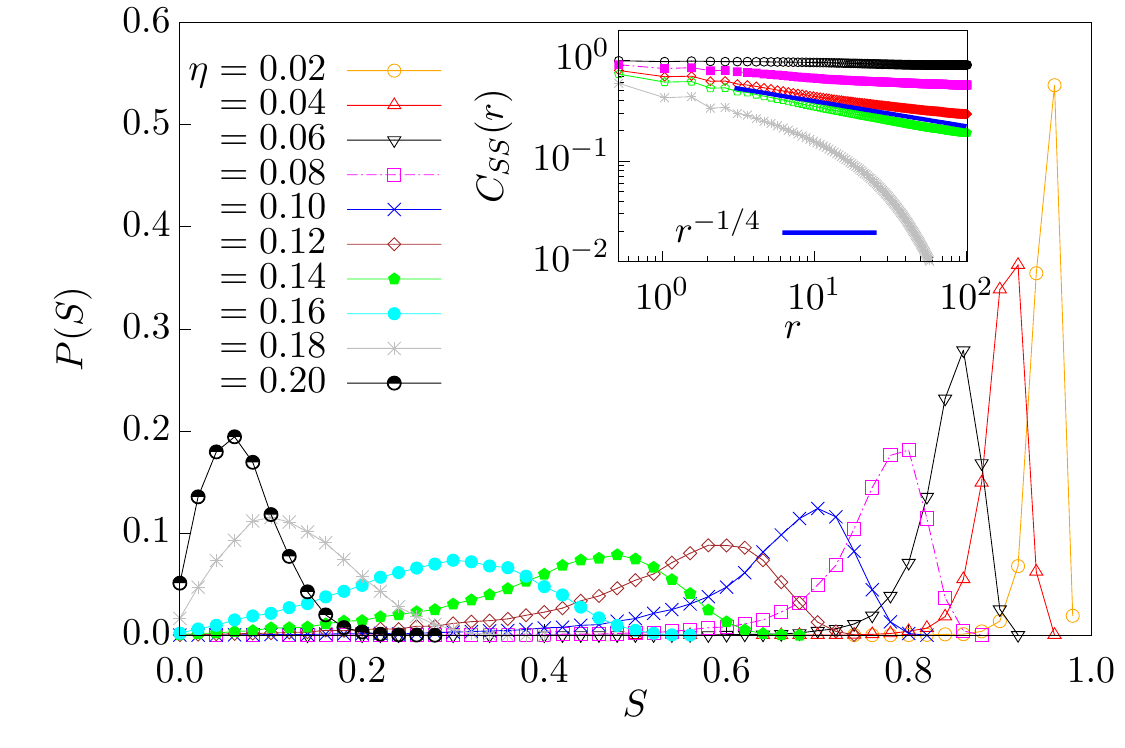} 
\caption{(color online) Probability distributions and correlation functions of the order parameter $S$ for system size  $N=16384$ and the mean density $\r=0.6$ at  $Pe=40$ and various orientational noise $\eta$.
With increasing noise, the distribution function remains unimodal, and the peak of the distribution shifts from high to low values. 
In the inset, the correlation function  $C_{ss}(r)$ is plotted at five noise strengths $\eta=0.02$ (Clustering phase), $0.08$(nematic phase), $0.12$ (just before the nematic-isotropic transition), $\eta=0.14$(just after the transition) and $0.18$ (isotropic). The blue line shows a $r^{-1/4}$ decay.
}
\label{NIPT3}
\end{figure}

%

The nature of the nematic-isotropic phase transition is further qualified using the probability distribution of the scalar order parameter, $P(S)$, across the transition. 
At a given activity,  
with increasing $\eta$ the probablity distribution remains unimodal, only the peak position shifts from $S \approx 1$ to values close to $S=0$~(Fig.~\ref{NIPT3}). This unimodal nature of the distribution functions signifies the absence of any metastable phase on the other side of the transition, a characteristic of continuous transitions. The distributions broaden near the phase transition point capturing the enhanced fluctuations.

The continuous nature of the transition is further characterized using the two-point correlation of the normalized nematic order parameter {$C_{ss}(r) =\la   \sum_{j,k} \cos[2(\h_j - \h_k)] \d(r - r_{jk})/\sum_{j,k} \d(r - r_{jk}) \ra$}. The correlations are plotted for different $\eta$ at $Pe=40$ in the inset in Fig.~\ref{NIPT3}. 
 Within the QLRO nematic phase, the correlations decay as power law $r^{-\nu}$, with the value of the exponent $\nu$ increasing with $\eta$. At the transition point the exponent $\nu \approx 1/4$, in approximate agreement with the prediction of equilibrium Kosterlitz-Thouless  transition~\cite{Kosterlitz1973}.

\begin{figure*}[!t]
\begin{center}
\includegraphics[width=0.85\linewidth]{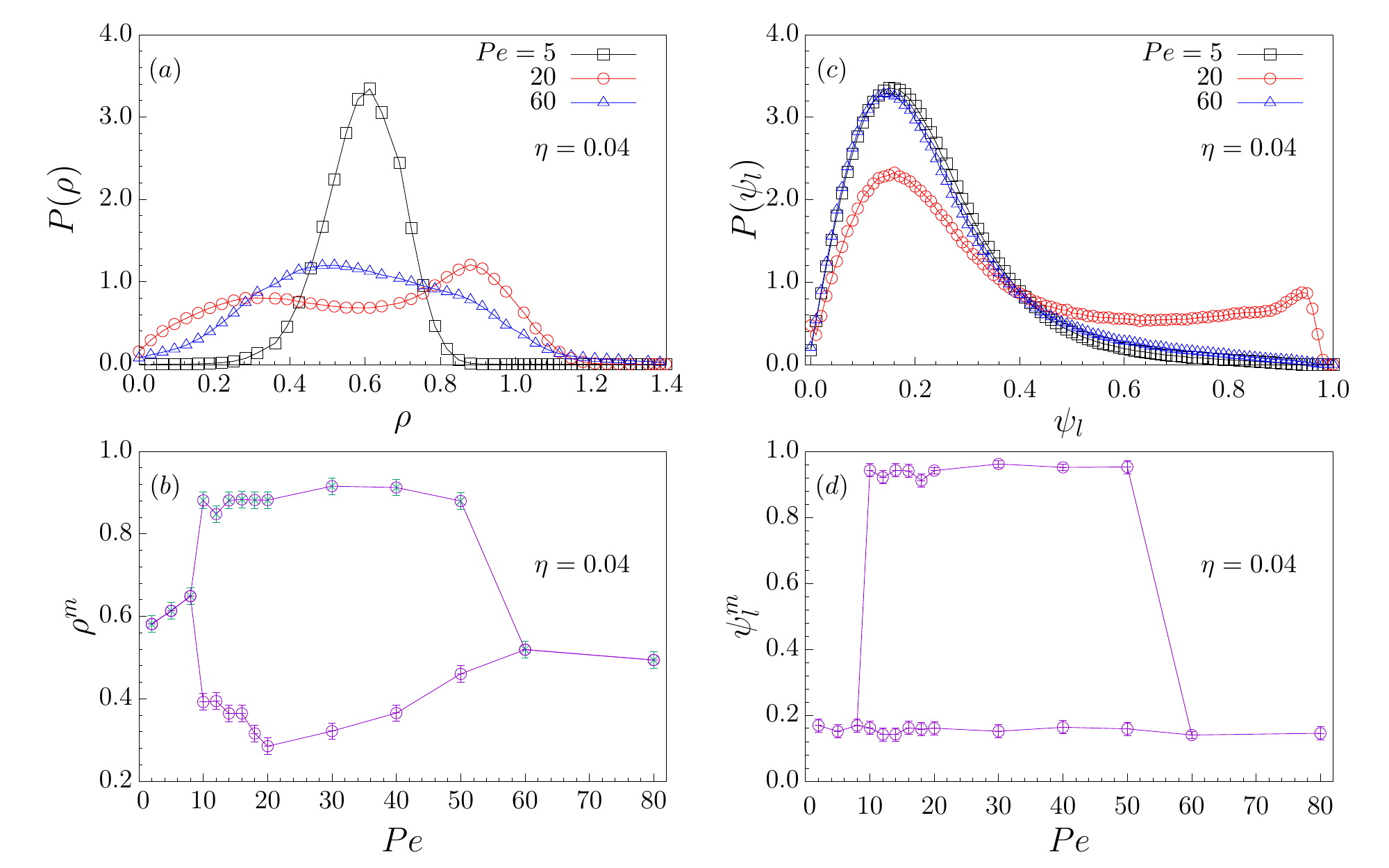}
\end{center}
\caption{(color online) ($a$) The density distribution $P(\r)$ and ($c$) the distribution of the local hexatic order $P(\psi_l)$ are shown at $Pe=5,\, 20,\,$ and $60$. These distributions are plotted at a fixed $\eta=0.04$, and different $Pe$ for a system size $N=16384$ and the mean density $\r=0.6$. The positions of the peaks in ($a$) and ($c$), denoted by $\r^m$ and $\psi_l^m$, are shown in ($b$) and ($d$) respectively as a function of $Pe$. Two peaks showing the phase coexistence lie  between $Pe=8$ and $Pe=60$. }
\label{peakPe}
\end{figure*}

Across the nematic-isotropic  transition, the  volume exclusion maintains a homogeneous density distribution. This behavior is quite unlike that observed for point particles, where increased  orientational disorder  leads to density inhomogeneity resulting in clustering, and band formation~\cite{Ginelli2010}. This remains the zero density or vanishing exclusion interaction limit of our model. At further smaller $\eta$, enhanced alignment reduces orientational diffusion, thereby, increasing persistence. As a result, as we show in the following section, the interplay of activity and volume exclusion leads to phase separation.

\subsection{Phase separation}
\label{sec:mips}

 The structural order in the coexisting phases are analyzed using a system of $N=16384$ particles in a volume of $L_x \times L_y$ with $L_x=157.37\,\s$ and $L_y=136.28\,\s$ fixing the dimensionless density to $\r=0.6$. The phase behavior is studied varying $Pe$ and $\eta$ values. At $\eta \lesssim 0.06$, the system undergoes phase separation with increasing $Pe$. We characterize the phase coexistence using the density distribution function $P(\r)$ and the distribution of local hexatic order $P(\psi_l)$,  where $\psi_l ({\bf r}) = |\vec \psi_l ({\bf r})|$ are calculated over local coarse-grained  volumes of $\ell_x \times \ell_y = 5.25\,\s \times 4.5\,\s$ around ${\bf r}$. 
The system is relaxed up to $10^{7}\delta t$, before analyses are  performed using $4000$ configurations  saved in intervals of $10^{4}\delta t$ obtained from simulations over further $4\times 10^{7}\delta t$.

\subsubsection{Re-entrant phase separation with $Pe$}
In the regime of small orientational noise $\eta \lesssim 0.05$, a re-entrant phase-separation  is observed with increasing $Pe$. 
A detailed characterization is presented in  Fig.\ref{peakPe} for a fixed $\eta=0.04$.  
At low activities, e.g., $Pe=5$, $P(\r)$ exhibits a single maximum at the mean density $\r= 0.6$ corresponding to the  homogeneous fluid~(see Fig.\ref{peakPe}($a$)\,). With increasing $Pe$, initially the distribution broadens, capturing the increased fluctuations in density. 
As the activity is further increased to $Pe=10$, clusters form, and the system phase separates into high and low density regions.
 
The phase separation is most pronounced at $Pe=20$.
The phase separation persists  up to $Pe=50$.
As activity is increased further to $Pe=60$, the high density clusters get destabilized and the bimodality disappears identifying remelting. 
The new unimodal distribution is significantly broad with respect to the fluid before the onset of phase separation at small $Pe$. It shows a shallow peak near $\r=0.48$~(Fig.\ref{peakPe}($a$)\,).  
This is due to the larger density fluctuations associated with the  stronger activity.  
The peak position(s) of the local density distributions for $\eta=0.04$ are plotted as a function of $Pe$ in Fig.\ref{peakPe}($b$). This clearly demonstrates phase coexistence over an intermediate regime of $Pe$, between $10$ to $50$.

\begin{figure*}[!t]
\begin{center}
\includegraphics[width=0.85\linewidth]{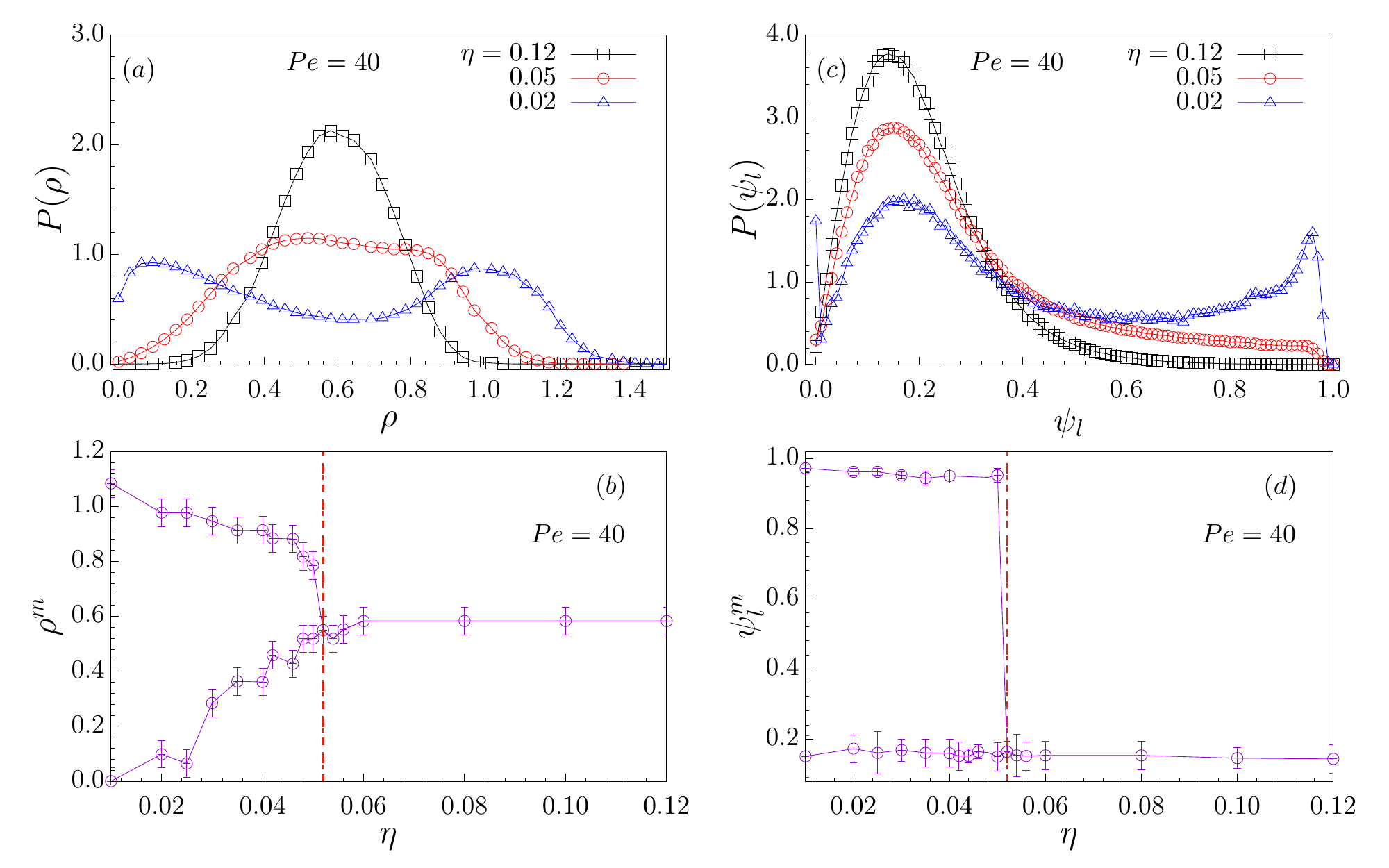}
\end{center}
\caption{(color online)  
For a fixed activity $Pe=40$, distribution functions  ($a$)~$P(\r)$ and ($c$)~$P(\psi_l)$ are shown at different noise strengths. The simulations are performed in a system size $N=16384$ and the mean density $\r=0.6$.  In ($b$) and ($d$) we plot the positions of the peaks of the distributions $P(\r)$ and $P(\psi_l)$, respectively. The vertical (red) dashed lines at $\eta=0.052$ in these two graphs indicate the boundary of phase separation.
}
\label{peakEta}
\end{figure*}

Associated with the clustering, a local hexatic order $\psi_l({\bf r})$ emerges. 
Three representative probability distributions $P(\psi_l)$ at $Pe$ values corresponding to the two homogeneous phases and the phase coexistence region, are shown in Fig.\ref{peakPe}($c$). 
At $Pe=5,\,60$ the distributions show a single maximum at a low hexatic order $\psi_l \approx  0.15$. 
At intermediate $Pe$ values, e.g., $Pe=20$, a bimodality in $P(\psi_l)$ is observed, showing emergence of a new large $\psi_l$ peak near $ \psi_l = 0.94 \pm 0.02$, associated with the emergence of the high density hexatic clusters. 
However, even within the phase coexistence region, $Pe=20$, the larger peaks of the  distributions remain at  a small $\psi_l$ value, reflecting the fact that the homogeneous fluid covers the larger spatial fraction of the system. 
The peak positions of the probability  distributions $P(\psi_l)$ are plotted in Fig.\ref{peakPe}($d$). 
This again captures the re-entrant phase separation, as  in  Fig.\ref{peakPe}($c$). With increasing $Pe$, the system shows a re-entrant transition  from a single fluid to a fluid-hexatic coexistence that finally returns to a single fluid.

\begin{figure}[!t]
\begin{center}
\includegraphics[width=0.9\linewidth]{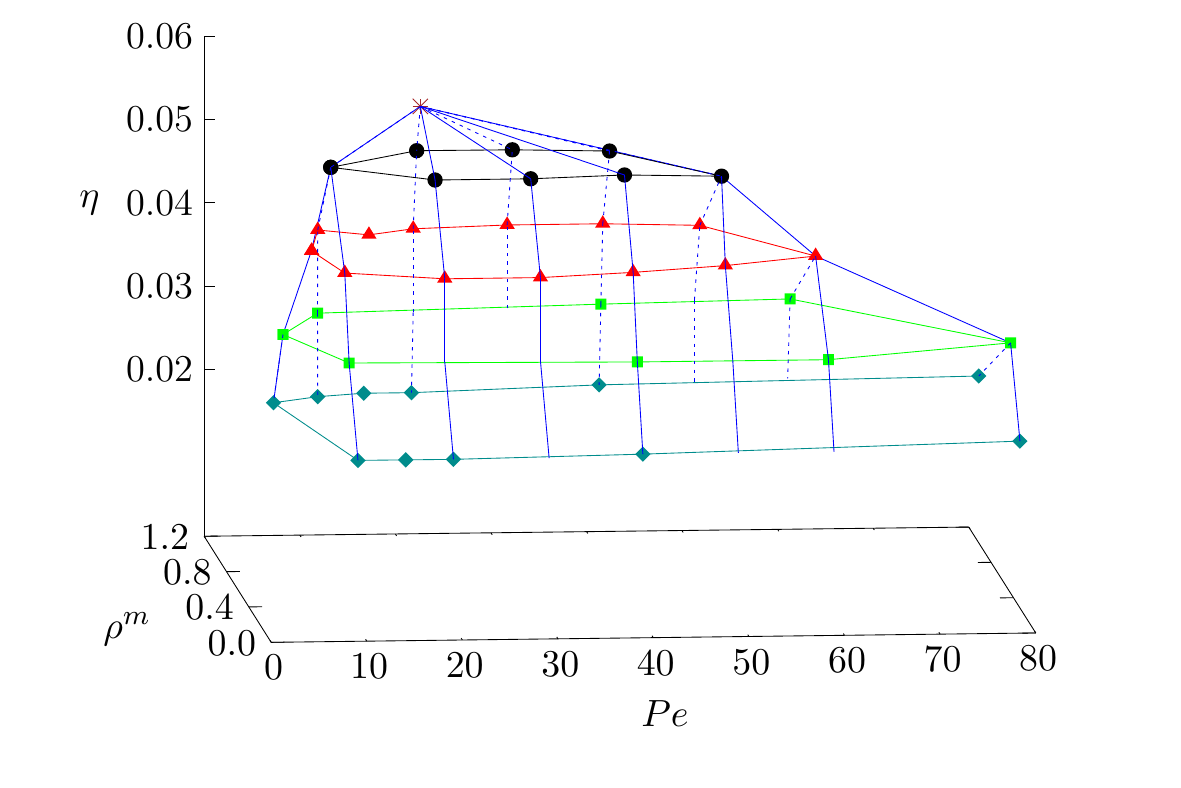}
\caption{(color online) The coexistence diagram in $Pe-\eta$, showing the binodal  $\r^m$ extracted from the peaks of local density distribution.
The points with various symbols specify binodals at fixed $\eta$. 
}
\label{fig:coexist}
\end{center}
\end{figure}
\begin{figure}[!t]
\begin{center}
\includegraphics[width=0.9\linewidth]{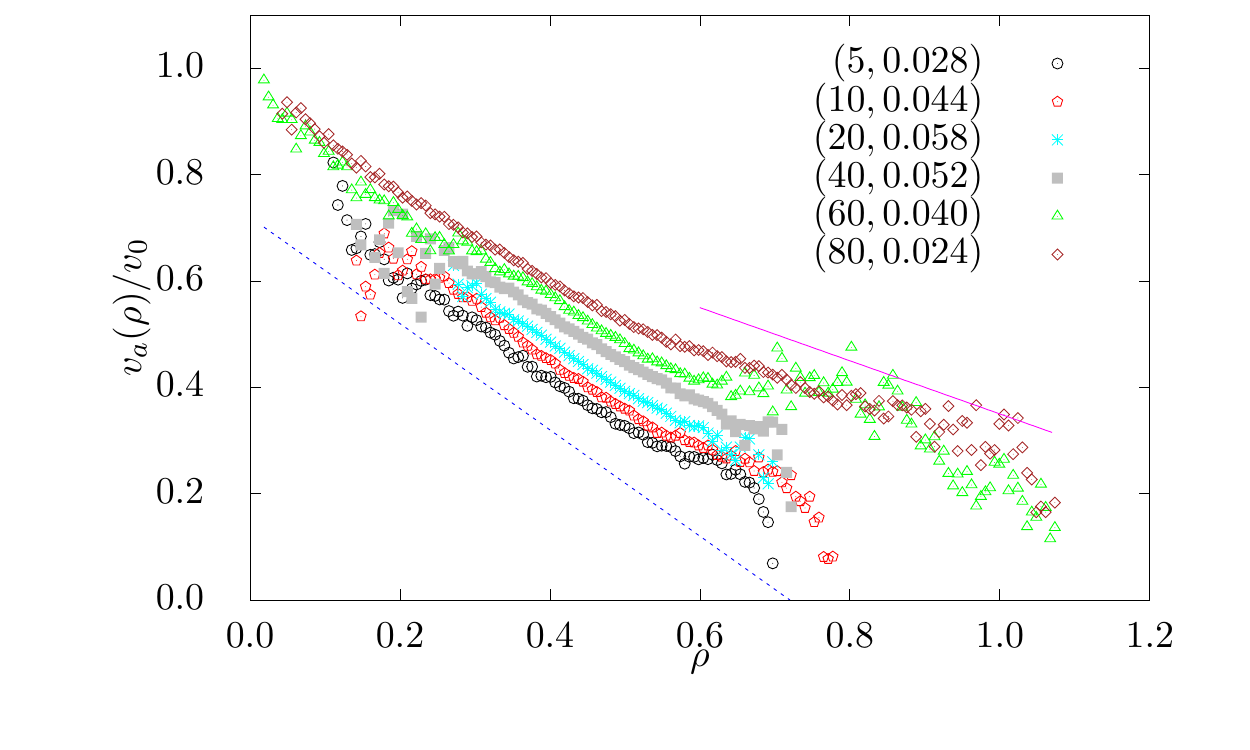}
\caption{(color online) The variation of the average local velocity $v_a(\rho)/v_0$ against the local density is shown along the phase separation boundary. The phase points  $(Pe,\eta)$ at which the data are calculated are indicated in the figure.
The solid and dashed lines denote linear decreases of the form $(a_0 - \r/2)$ and $(b_0 -  \r)$, respectively.
}
\label{LOCVEL-RHO}
\end{center}
\end{figure}

%
\begin{figure*}[!t]
\begin{center}
\includegraphics[width=0.95\linewidth]{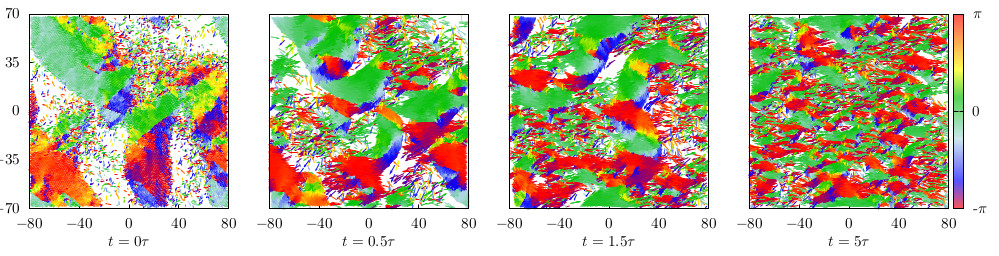}
\caption{(color online) The destabilization of clusters formed at $\eta=0.04$ and $Pe=10$ ($t=0\,\t$) are shown after quenching to $Pe=70$, keeping $\eta$ fixed. The velocity vectors (small arrows) are shown with the  color coded velocity directions with respect to the $x$-axis. 
}
\label{dest}
\end{center}
\end{figure*}

\subsubsection{Phase separation with $\eta$}
To demonstrate the dependence of the phase separation on the strength of orientational fluctuations $\eta$, we consider the system at a sufficiently high activity $Pe=40$ and vary $\eta$.
As we have seen above, at this activity the system remains phase separated at $\eta=0.04$. 
In Fig.\ref{peakEta}($a$) we show the variation of $P(\r)$ with $\eta$. 
At low $\eta$ ($=0.02$, blue up triangles), the system remains phase separated,  captured by the bimodality of $P(\r)$. With increasing $\eta$, the peaks in the distribution slowly come together and merge at $\eta=0.05$ (red circles). 
For larger $\eta$, the density distribution remains unimodal (e.g., $\eta=0.12$, black squares) with the peak at the mean density $\r=0.6$, signifying a homogeneous fluid phase. 
The peak positions of $P(\r)$ are plotted against $\eta$ in Fig.\ref{peakEta}($b$). This shows how the coexistence lines merge together as a function of $\eta$ at a critical point. 
This behavior is similar to the merger of liquid-gas coexistence lines as the critical temperature is approached from below.
The detailed properties of such a critical point has been explored for the ABPs recently~\cite{Siebert2018}, and remains to be studied in the current context in future.  
The distributions of the local hexatic order $P(\psi_l)$ show pronounced multi-modality at the phase coexistence. 
The peak near $ \psi_l =  0.96$ captures the significantly large hexatic order of the active clusters. 
The other peak near $\psi_l =  0.17$ signifies the coexisting homogenoeus fluid. In addition, we observe a peak at $\psi_l=0$, which corresponds to the voids generated at the cost of clustering. The non-zero peaks of $P(\psi_l)$ are plotted against $\eta$ in  Fig.\ref{peakEta}($d$), recapturing the loss of phase coexistence with increasing $\eta$, as in Fig.\ref{peakEta}($b$).

Using the above analysis it is possible to construct a combined phase- coexistence diagram in the $Pe-\eta$ plane as is shown in Fig.\ref{fig:coexist}. The $\r_m$ values denote the binodals
extracted from the peaks of the probability distribution of the local density. The apex of the plot at large $\eta=0.056$ denotes the critical point below which phase coexistence becomes possible. 
The lines indicate the boundary of the onset of phase coexistence. As is clear from the plot, the coexisting densities separate out more strongly at lower $\eta$. At a given $\eta$, after the onset  
of active phase separation appears at a low $Pe$, coexisting densities first separate out to finally merge as $Pe$ is increased. At lower $\eta$, large activity supports formation of flocks. As a result
this merger happens at a larger $Pe$. Moreover, lowering of $\eta$ reduces the orientational diffusion supporting MIPS at small $Pe$, thereby reducing the $Pe$ value at which phase separation appears. 
The flocking and MIPS mechanisms are discussed further in the following section.

\subsubsection{Cluster formation}

In the absence of alignment, volume exclusion  induces slow down of active motion with increasing local density generating a positive feedback mechanism that mediates MIPS. 
However, the clustering we observe is deep within an orientationally ordered nematic phase. 
This raises  the question whether the MIPS- like mechanism still plays any role in clustering, or the phase coexistence is due to the onset of flocking related to orientational fluctuations.
It was argued before that alignment for polar disks reduces rotational diffusion, and thereby may favor MIPS~\cite{Martin-Gomez2018, Sese-Sansa2018}. On the other hand, steric alignment of self propelled rods beyond an asymmetry parameter produces  destabilizing torques to break MIPS clusters~\cite{Grossman2019, Shi2018, Weitz2015}. 
We note that, for our self propelled disks in the nematic phase with  antiparallel active orientations, crowding effects may slow down the propulsion speed. 
In fact, as we show below, this is the dominant mechanism of clustering in the small $Pe$ regime. 

Here we compute the  self-propulsion speed $v_a(\r)$ reduced by the inter-particle repulsion at a local density $\r$. We proceed by calculating the coarse- grained quantity  $(1/n_r)\mid \sum_{i\in r}^{n_r} [v_0 {\bf n}_i - \mu \nabla_i U(r_{ij})\,]\cdot {\bf n}_i \mid$ in cells of volume $\ell_x \times \ell_y$ having $n_r$ particles~\cite{Sese-Sansa2018}, such that the instantaneous density is $(\pi/4)n_r \s^2/\ell_x \ell_y$. 
Within the MIPS paradigm,  it is expected that  $v_a(\r) = v_0  (1- a \r)$ where $a$ depends on $v_0$, a stall time at collision, and a scattering cross-section~\cite{Stenhammar2013}. 
Fig.\ref{LOCVEL-RHO} shows the variation of $v_a(\r)$ along the phase separation boundary. 
In the small $Pe$ regime, between $Pe=2$ to $20$,  $v_a(\r)$ decreases linearly with $\r$ with a single slope $a \approx 1$, in agreement with the MIPS expectation.  
However, at higher $Pe$, the behavior starts to get dominated by flocking, shown by the slow down of the decay of $v_a(\r)$ at higher $\r$. 
Note the cross-over to a second slope $a \approx 1/2$ at large $\r$ for $Pe > 20$. 
This happens as a cluster starts to get an effective polar alignment such that the active velocities of particles do not completely block each other. Similar cross-overs have been observed earlier for ferromagnetically aligning ABPs~\cite{Sese-Sansa2018}.

\begin{figure*}[!t]
\begin{center}
\includegraphics[width=0.9\linewidth]{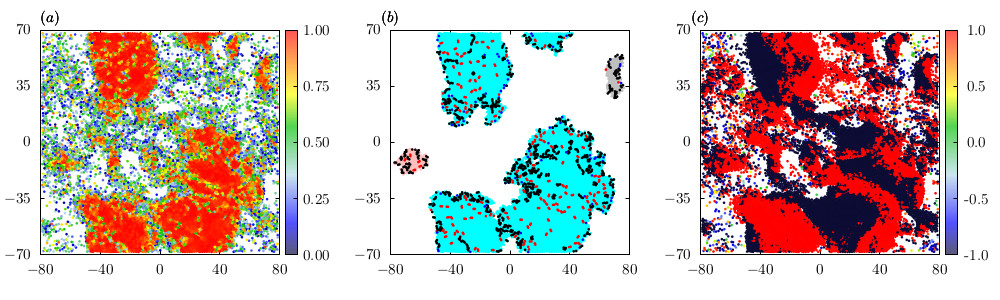}
\caption{(color online) 
Structures at $Pe=40$ and $\eta=0.02$. ($a$)~The hexatic order associated with the all $k$ particles $\mid \vec{\psi}_6^k \mid^2$. ($b$)~The three largest clusters are shown: the largest with cyan background, the second largest with grey background and the third largest with pink background. The topological defects, dislocations~(red), disclinations~(blue) and the grain boundaries~(black) within these three clusters are pointed out. $(c)$~The same  configuration as in ($a$) is plotted with color codes denoting the projection of the active orientation of particles along the overall nematic orientation, $\hat{n}_k\cdot {\hat s}$. }
\label{CONF40}
\end{center}
\end{figure*}

At small $Pe$, colliding particles slow down as in MIPS. The size of a hexatic cluster is maintained by the absorption of particles from the surrounding fluid having density $\r_l$, with a rate $r_{\rm on} = \r_l v_0/\pi$ and evaporation from the cluster surface due to reorientation of the direction of self-propulsion with a rate $r_{\rm off} = \k D_\h (\eta)/\s$, where $\k$ denotes the total number of particles leaving the cluster in each escape  ~\cite{Redner2013}. 
The orientational relaxation is controlled by the noise strength $\eta$, such that the related rotational diffusion constant $D_\h \sim \eta^2$~(see Appendix-\ref{sec_app:C}).  
The mean density of the system in the phase coexistence regime can be expressed as  $\r = f \r_h + (1-f) \r_l$, where $\r_h$ ($\r_l$)  denotes the mean  density of the hexatic (liquid) fraction, and $0 \leq f < 1$ denotes the amount of hexatic fraction. At the onset of phase coexistence, $f=0$ and $\r = \r_l$. Thus, equating $k_{\rm on} = k_{\rm off}$ and setting $\r_l = \r$, we get 
$v_0 = \pi \k D_\h(\eta)/\s \r$. 
This leads to the condition $Pe = v_0 \s/D = \pi \k D_\h(\eta)/\r D$. Using $D_\h \sim \eta^2$ we obtain the  prediction for MIPS- like phase boundary $\eta \sim Pe^{1/2}$, plotted in the phase diagram Fig.\ref{PD}($a$). This  captures the behavior of the low $Pe$ boundary for the onset of active phase separation.

To illustrate the melting process of hexatic clusters at high $Pe$, in Fig.\ref{dest}, we show the evolution of a steady state  configuration after quench from within the phase coexistence regime at $\eta=0.04$ and $Pe=10$ to $Pe=70$, an activity beyond the remelting boundary. The plots show velocity orientation (color code) and speed (arrow) of particles, e.g., right moving particles (orientation $0$) are denoted by green and the left moving ones by red (orientation $\pm\pi$). If counter moving polar patches do not block each other, they slide past. In other places, e.g., in the largest cluster in the  upper- left corner at the initial configuration $t=0\, \t$~(Fig.\ref{dest}), a left moving patch creates channel and squeezes  through the majority right moving particles. 
The polar patches eventually detach, destabilizing the big clusters to smaller clusters. 
Similar ejection of polar clusters were observed before in self propelled rods~\cite{Weitz2015}.
The ejection process continues to finally melt the hexatic clusters, e.g., see the last configuration after a time lapse of $5\t$ in Fig.\ref{dest}. The detailed dynamics can be observed in the movie added in the  supplemental material~\footnote{The video shows the time evolution of the local density and velocity fields over $5\,\t$ after a quench from $Pe=10$ to $Pe=70$ at $\eta=0.04$. The color code in the video follows that in Fig.\ref{dest}.}.

\begin{figure}[!h]
\includegraphics[width=0.95\linewidth]{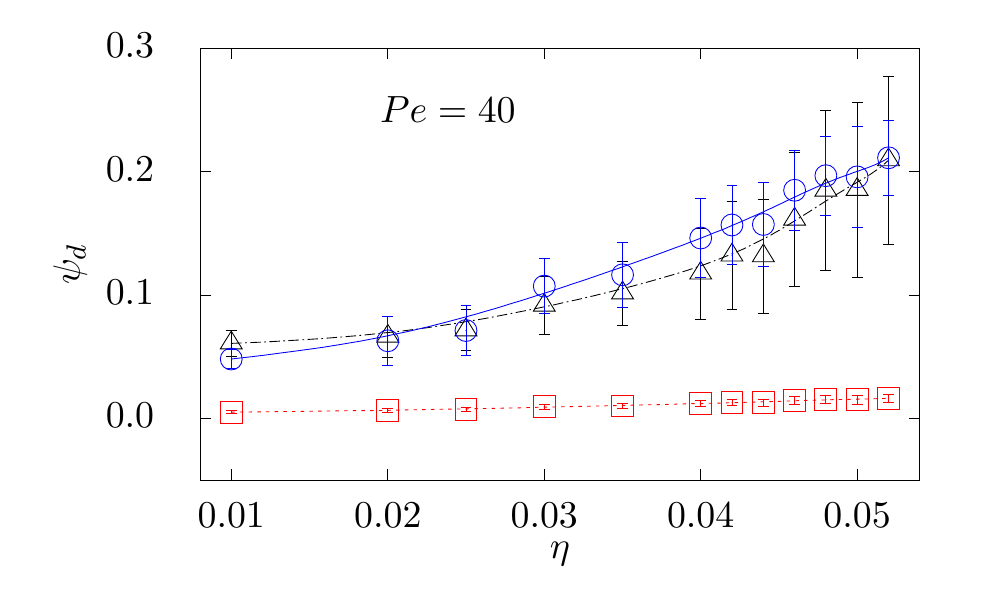}
\caption{(color online) 
The variation of the fraction of dislocations (black up triangle), grain boundaries (blue circle) and disclinations (red square) with $\eta$ across the clustering transition are shown for $Pe=40$. The lines are guide to the eye.
}
\label{DFFrac}
\end{figure}

\subsection{Topological defects}
In this section we investigate the role of the topological defects in the melting of the active hexatic clusters. We define the coordination number of each particle $n_v$ as the number of Voronoi neighbours it has. The $n_v \neq 6$ particles  constitute the topological defects. 
Within hexatic clusters, particles with  $n_v=5$ and $n_v=7$ stay as bound pairs. 
Such $5-7$ bound pairs, well separated from each other, identify the presence of dislocations in hexatic. 
For example, see the upper left inset in Fig.\ref{PD}($a$) displaying several  $5-7$ bound pairs in a portion of a hexatic cluster of the active system. 
Unbinding of such $5-7$ pairs, generate free disclinations. As we show, interconnected chains of $5-7-5-7..$ defects form grain boundaries, held by local polar patches of the hexatic clusters. 

We analyze the variation of the fraction of particles belonging to different defect types within the hexatic clusters, with increasing $\eta$ at $Pe=40$~\cite{Qi2014}.  A typical configuration at $\eta=0.02$ is shown in Fig.\ref{CONF40}. 
In Fig.\ref{CONF40}($a$) we display the hexatic order associated with all the $k$ particles $\mid \vec{\psi}_6^k\mid^2$. The large clusters have an  overall large hexatic order $\mid \vec{\psi}_6^k\mid^2 \approx 1$.  The local drops in the values of $\mid \vec{\psi}_6^k\mid^2$  within the large clusters are associated with formation of topological defects.  
In Fig.\ref{CONF40}($b$) we show the three different defect types separately --  dislocations (red points),  disclinations (blue points) and grain boundaries (string of black points) within the three largest clusters colored with cyan (largest cluster), grey (2nd largest cluster) and pink (3rd largest cluster). 
To understand the localization of the grain boundaries, we display the projection of the local polar orientation $\hat n_k = (\cos \h_k, \sin \h_k)$ of the particles onto the orientation of instantaneous overall nematic order ${\hat s} = {\bf s}/\mid {\bf s} \mid$ in Fig.\ref{CONF40}($c$). A comparison between Fig.\ref{CONF40}($c$) with ($b$) shows that the grain boundaries appear at the interface of anti-parallel polar patches of activity. At high activity, such as $Pe=40$, the opposing active push from these neighboring polar regions leads to the formation of the grain boundaries within the clusters.

To quantify the behaviour of the topological defects across the melting transition, we compute the fraction $\psi_d$ of particles associated with the dislocations, grain boundaries and disclinations within the largest cluster of a given configuration. 
The averaging is performed over more than $3000$ well separated configurations in the steady state. In Fig.\ref{DFFrac} we show the variation of these three average quantities with increasing $\eta$, at $Pe=40$.  In the active clusters, all the disclinations are thrown out to the hexatic- fluid interfaces. The fraction of the disclination within the hexatic cluster, thus, remains largely unchanged. The fraction of dislocations increases with $\eta$, as well as the fraction of grain boundaries. Note that the dislocations characterize hexatic phase, and do not destabilize it. Thus the melting of the active hexatic at large $Pe$ is mediated by grain boundary formation along the interface of oppositely aligned polar patches within a cluster.

\subsection{Cluster size and dynamics}
\label{sec:structures}
\begin{figure}[t]
\begin{center}
\includegraphics[width=8cm]{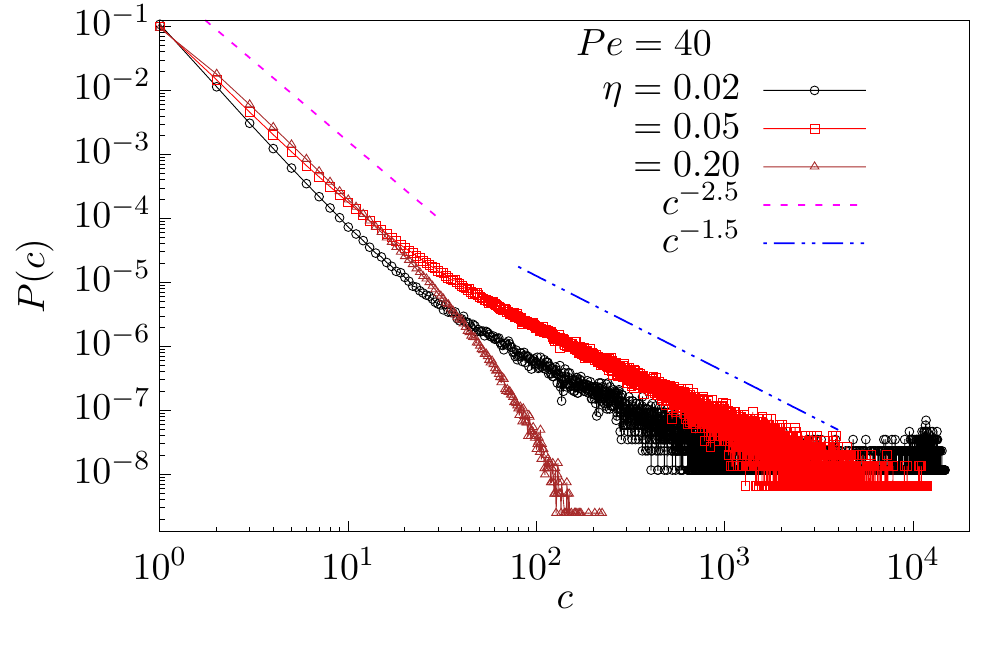}
\end{center}
\caption{(color online) 
The steady state cluster size distribution is shown for three different orientational noise strengths $\eta=0.02$\,(within the  phase separated state), $0.05$\,(around the onset of phase separation) and $0.20$\,(in the  isotropic fluid phase). All of them show an approximate power law decay  $P(c)\sim c^{-2.5}$ for smaller clusters. For large $c$, this crosses over to  an exponential decay in the isotropic phase. Within the region of phase separation, the distribution crosses over to a longer power law tail $\sim c^{-1.5}$, in addition to showing clusters of large masses having small but uniform weight. The simulations are performed in a system size $N=16384$ and the mean density $\r=0.6$.  
}
\label{CLD1}
\end{figure}

As has been mentioned before, the coexistence at the active phase separation is characterized by the presence of large correlated clusters. They are identified using a clustering algorithm, defining particles within a separation $\s$ to be part of a single cluster~\cite{Allen1989}. The number of particles belonging to a cluster determines the cluster size $c$. Three  typical steady state cluster size distributions $P(c)$ in the homogeneous fluid phase and at the active phase coexistence  are shown in Fig.\ref{CLD1}. 

\begin{figure}[t]
\includegraphics[width=0.95\linewidth]{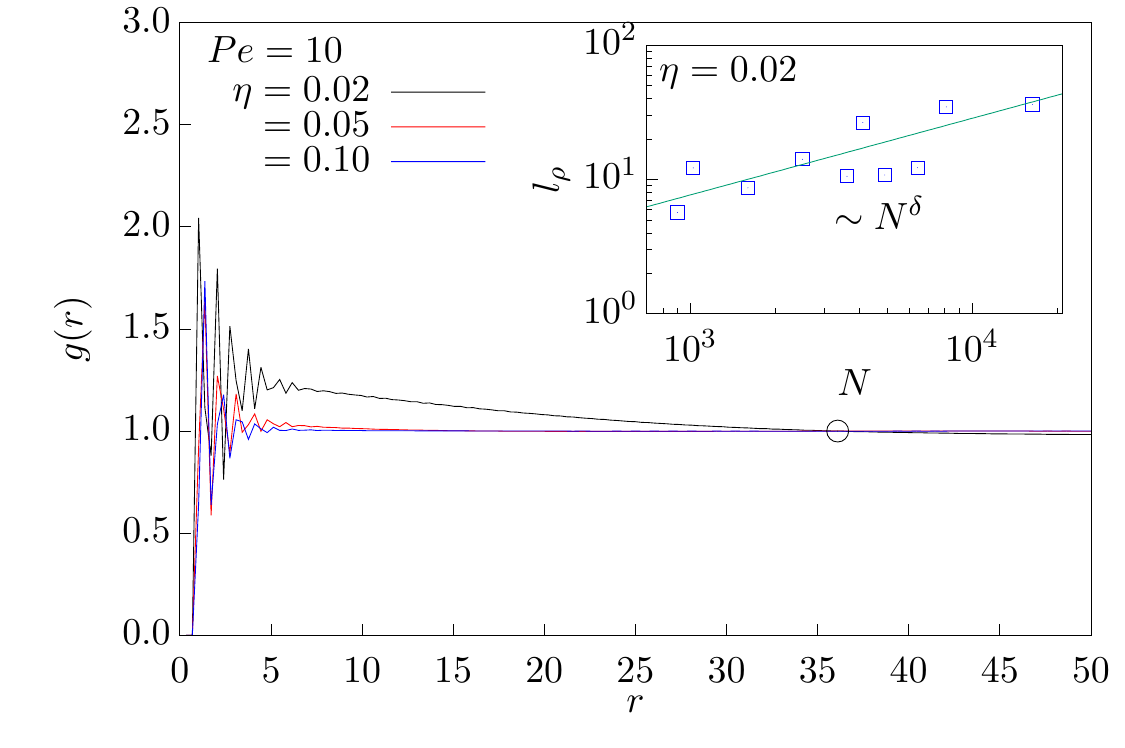}
\caption{(color online) The density-density correlation function  $g(r)=\langle \rho(0) \rho(r)\rangle / \la \r \ra^2$ as a function of the separation $r$ at various $\eta$, calculated for a system of $N=16384$ particles at the mean density $\r=0.6$ and activity $Pe=10$. The correlation length $l_\r$  is given by the $r$- value at which $g(r)$ approaches unity, and at $\eta=0.02$ is  indicated by the  $\circ$ in the figure. The inset shows that the correlation length remains a finite fraction of the system size, $l_\r/L_0$ does not decay with $N$. }
\label{GrCT}
\end{figure}

\begin{figure*}[t]
\begin{center}
\includegraphics[width=0.95\linewidth]{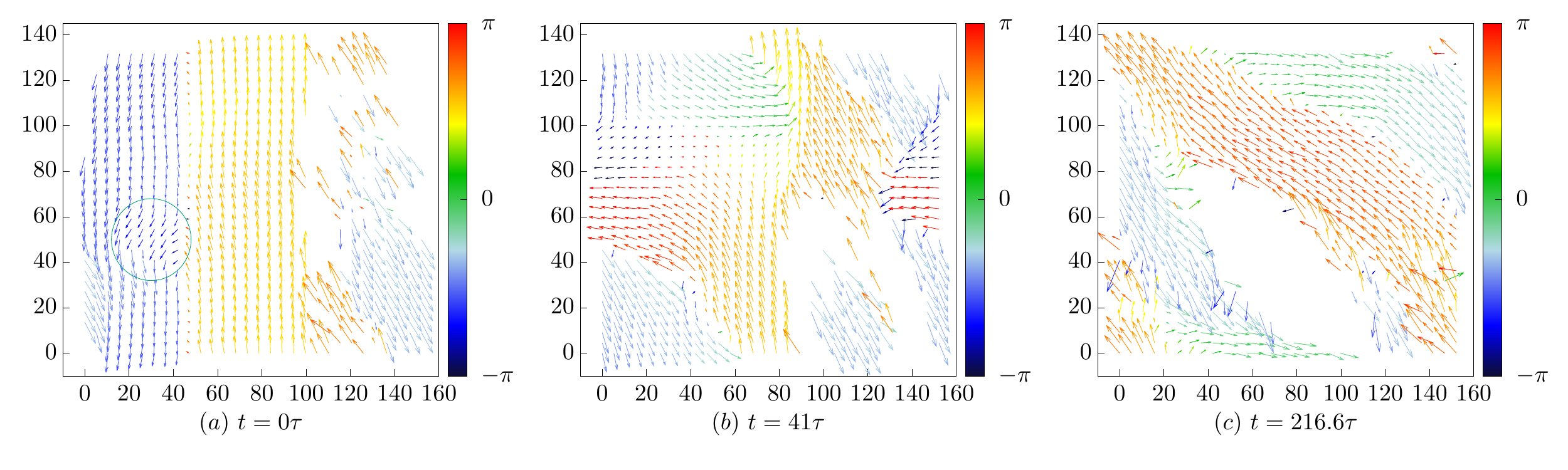}
\caption{
(color online) The structural and dynamical evolution of the system  at $Pe=10$ and $\eta=0.0002$. 
In $(a)-(c)$ we show the local velocity fields of the system  at well separated time instances.  The arrows show the local velocity vectors. The directions of the velocity field are  further highlighted by the color code. $(a)$ A steady state configuration at a given time $t=0 \,\tau$. The velocity fields show that  anti-aligned domains glide past each other. $(b)$ At a later time $t=41\, \tau$, the lane formation starts, and $(c)$ at $t=216.6\, \tau$ well formed lanes are observed.
}
\label{CONF}
\end{center}
\end{figure*}

The distribution of small clusters are approximately independent of the phase, and the weight decays as $c^{-2.5}$. The decay of the  distribution for large clusters, on the other hand, changes qualitatively with the change in phase.  
In the homogeneous fluid phase, it decays exponentially. 
In contrast, in the regime of phase separation, the tail of the distribution shows a power law decay $P(c)\sim c^{-1.5}$. This appears in addition to a small but approximately uniform weight associated with the  largest clusters. 
The change of the tail of the  cluster size distribution from exponential decay to a power law is qualitatively similar to that observed in the conserved mass  aggregation model (CMAM)~\cite{Majumdar1998}, however, the exponents in our case differ significantly. 
Moreover, for the largest clusters structural ordering competes with ejection of polar fractions, not allowing stabilization of infinite-aggregates, unlike in the CMAM.

A measure of the spatial extension of the clusters is the correlation length obtained from the two point density correlation function  $g(r)=\langle \rho(0) \rho(r)\rangle / \la \r \ra^2$. 
In Fig.\ref{GrCT} the variation of $g(r)$ is shown in the fluid and the phase coexistence region. 
The typical correlation length, $l_{\rho}$, is estimated as the value of  the separation $r$ beyond which the correlation decays to $1 \pm 0.01$. 
As is shown in Fig.\ref{GrCT}, in the  regime of phase separation ($Pe=10,\, \eta=0.02$) one gets a large $l_\r$~($\approx 35\,\s$) corresponding to the large extension of the clusters. 
However, in the homogeneous fluid phase the correlation length becomes small ($l_\r \approx 5\s$). 
If the appearance of phase separation is not a finite size effect, the cluster size $l_\r$ should remain a finite fraction of the system size $L_0=(L_x \times L_y)^{1/2}$, in the thermodynamic limit of infinite $N$. To test this we perform a system size scaling. As we show in the inset of Fig.\ref{GrCT}, indeed, the ratio $l_\r/L_0$ does not decrease with system size $N$.


The collective motion is dominated by flock- like behavior at large $Pe$ as well as small $\eta$, at which alignment dominates. This consists of (i)~structurally ordered stripes sliding past each other, (ii)~jamming of polar  clusters, (iii)~which eventually break via formation of counter-propagating channels. 
The behaviors are shown in  Fig.\ref{CONF}($a$)-($c$) with the help of local vector fields identifying the orientations and amplitudes of the locally coarse grained velocity of particles. The coarse- graining is performed over cells of size $\ell_x \times \ell_y$, and the velocity fields are calculated from displacements obtained over $0.1\,\t$. 
The orientations of velocity are further highlighted by color codes explained in the figure. 
The snapshots are shown at three different well separated time instances in the steady state, deep inside the phase coexistence regime. 
For the ease of illustration, we have chosen $Pe=10$ and $\eta=2 \times 10^{-4}$, as the clusters at such a low $\eta$ are system spanning in extension and have long life-times.        
In  Fig.\ref{CONF}($a$) we plot the first steady-state snapshot of local velocity fields. It shows two large counter propagating clusters that glide past each other, in the up and down directions~($\approx \pm \pi/2$).  
The velocity field at the interface of the two clusters vanishes, before changing direction.  
With time, the velocity field from one domain may penetrate the other due to dynamical fluctuations along the interface. 
In Fig.\ref{CONF}($a$), such events are identified by a bulge of vectors marked within a circle. This generates instability towards  the formation of lanes. At a later time, $t = 41 \t$, as is shown in  Fig.\ref{CONF}($b$), particles in this region starts to flow leftward (direction $\approx \pm \pi$).     
They create path for other particles to trickle along. 
As time progresses, this local trickling takes the form of a stream of particles forming counter propagating lanes, as is shown in Fig.\ref{CONF}($c$) at $t=216.6\,\t$. 
The crossovers between these different regimes happen faster at higher $Pe$ and $\eta$. 

\section{Discussion}
\label{sec:discussion}

{We have studied the combined effect of volume exclusion and nematic alignment on collective behavior of active polar particles. Using numerical simulations, we have presented a phase diagram encapsulating the structural and orientational transitions with changing activity $Pe$ and noise $\eta$. 
In the equilibrium limit of vanishing activity our system behaves as a high density fluid. On the other hand, the limit of vanishing interaction reduces it to the nematically aligning point particles~\cite{Ginelli2010}.  
The nematic-isotropic transition in activity obtained by increasing $\eta$ does not change the homogeneous nature of the fluid phase of the excluded volume particles. 
This behavior is in striking contrast to the properties observed for point particles~\cite{Ginelli2010}, where the nematic-isotropic transition was accompanied by phase separation.
The volume exclusion in the current model, in addition to a significant rotational diffusivity, $D_\h \sim \eta^2$, even at the nematic-isotropic transition, suppresses such fluctuations. 

However, we do find clustering transition in our system, but at a lower strength of $\eta$ associated with larger persistence, $D_\h^{-1}$. As we have shown, the competition of persistence and volume exclusion led to a kinetic slowdown with density. This provided a MIPS mechanism towards phase separation at small $Pe$.  
The nematic fluid undergoes a transition from the homogeneous phase to an active phase separation characterized by the formation of high density hexatic clusters.    
We found a re-entrant fluid- phase coexistence- fluid transition with increasing $Pe$.
At the small $Pe$ regime, volume exclusion dominates -- the formation of clusters is mediated by the slow-down of colliding particles providing a positive feedback, as in MIPS. The phase boundary in this regime follows a form $\eta \sim Pe^{1/2}$, that could be  motivated using the MIPS phenomenology.  

With increasing $Pe$, the opposing push from the counter propagating polar sub-regions of a nematic cluster leads to formation of grain boundaries. Once the active push gets large enough, opposing polar regions squeeze past each other forming lanes. 
At higher $Pe$, the clusters eject polar patches that form co-moving flocks. At high enough $Pe$, this breaking and ejection of polar patches continue to dissolve  larger clusters into smaller ones, finally leading to remelting. As in the high $Pe$ regime, at the lowest orientational noise strengths,  alignment dominates leading to flock- like collective motion that consists of    sliding clusters, jamming, and lane formation. 

An experiment on active colloidal disks, recently, examined the combined impact of volume exclusion and alignment in the phase separation kinetics at high activity~\cite{Linden2019}. Similar experiments, may potentially verify our predictions.    
A coarse grained hydrodynamic description combining active alignment and density dependent slowdown remains to be developed, and may provide further insight into the phase transitions of such systems. 
}

\section*{Acknowledgements}
The simulations were performed on SAMKHYA, the high performance computing facility at IOP, Bhubaneswar. DC thanks ICTS-TIFR, Bangalore for an associateship, and SERB, India, for financial support through grant number EMR/2016/001454.

\appendix


\section{Quasi long range order}
\label{sec_app:B}
\begin{figure}[!t]
\begin{center}
\includegraphics[width=0.95\linewidth]{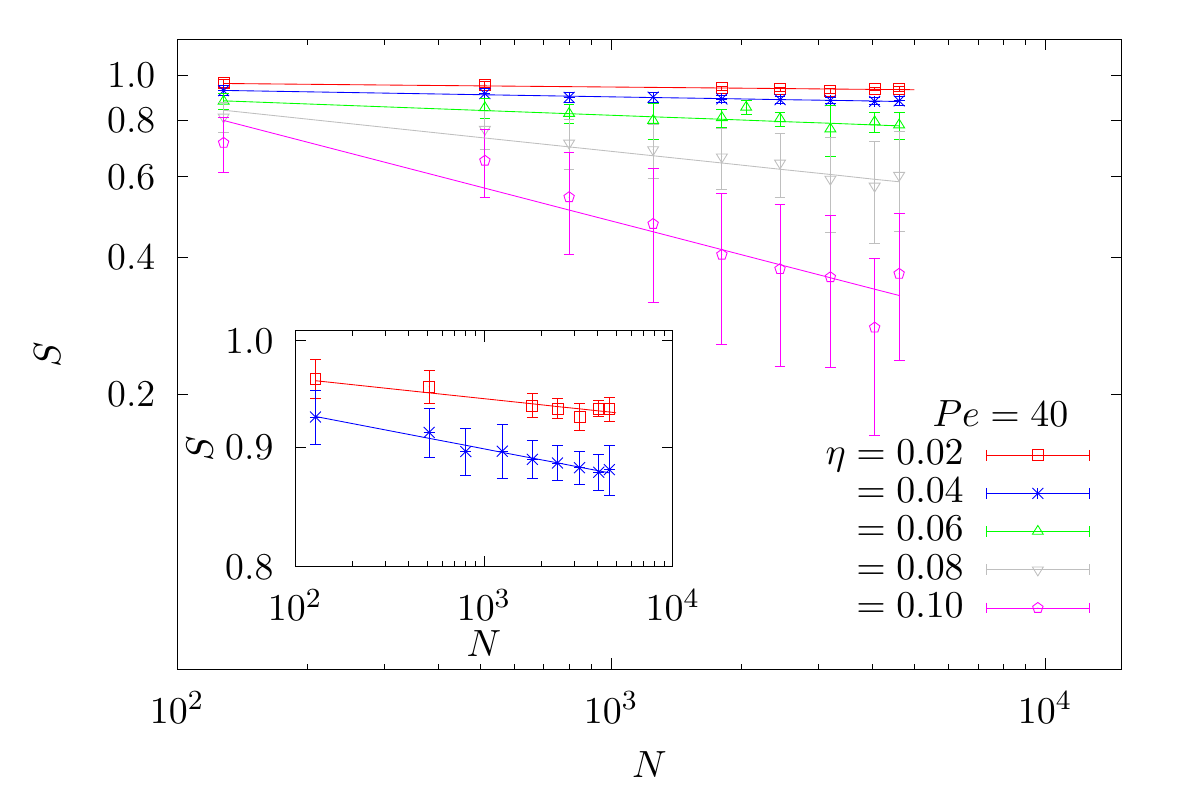}
\end{center}
\caption{(color online)
We show the dependence of the order parameter $S$ against the system size $N$ for different $\eta$ values within the nematic phase, at $Pe=40$. An algebraic decay of the form $S =  S_0(\eta)\, N^{-\a(\eta)}$ is observed for all $\eta$. The decay exponent increases as one approaches the transition. The inset magnifies the algebraic decay for the smallest two $\eta$ values. 
}
\label{QLRO}
\end{figure}

Here we demonstrate the nature of the nemtaic order before the nematic-isotropic transition. To this end, we perform a system size analysis by calculating the scalar order parameter $S$ for system sizes $N=128$ to $9216$, at a fixed  $Pe=40$, and varying $\eta=0.02-0.10$~(Fig.\ref{QLRO}). 
With increasing $N$, we find an algebraic decay $S=S_0\, N^{-\a}$, where both $S_0$ and the exponent $\a$ depend on $\eta$.  Even at the smallest $\eta$, the algebraic decay of the order parameter with system size is evident~(see inset of Fig.\ref{QLRO}). This is a signature of the quasi- long ranged order (QLRO) of the active nematic phase mentioned in Sec.\ref{sec:iso_nema}. The exponent $\a$ increases with $\eta$.

\section{Scaling of the $S$-$\eta$ curves}
\label{sec_app:A}

\begin{figure}[!t]
\includegraphics[width=0.95\linewidth]{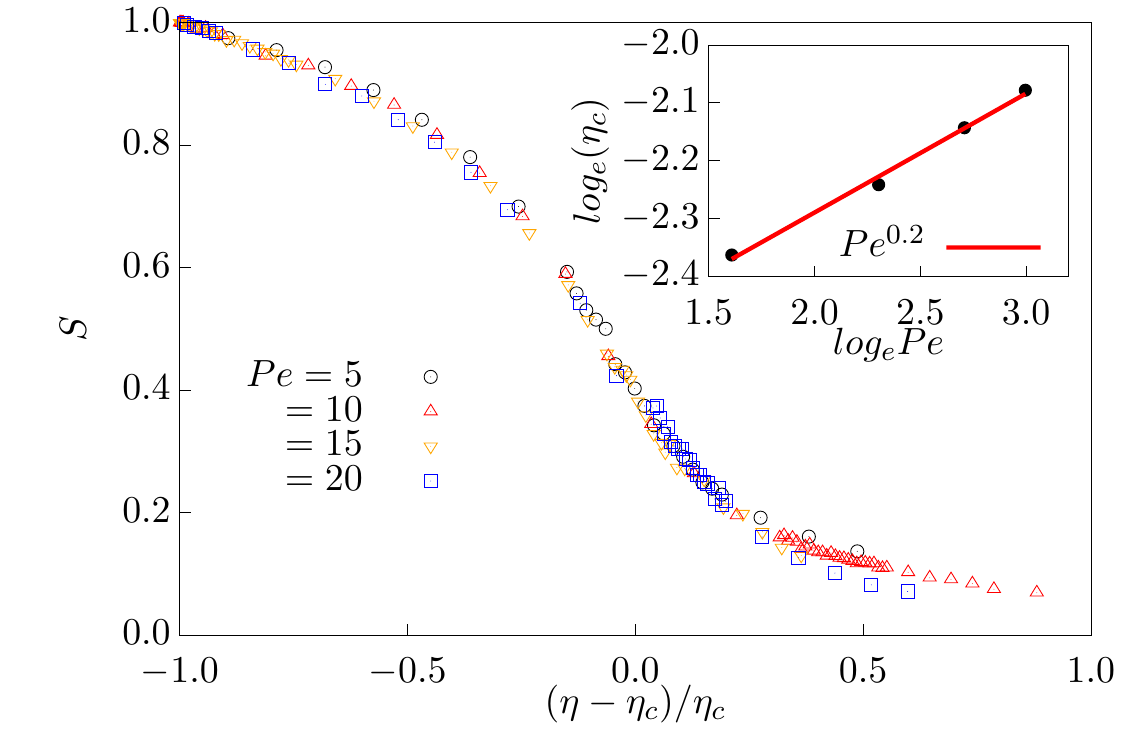}
\caption{(color online)
Scaling of the S-$\eta$ curves at  four different $Pe$ values. Transformation of the $\eta$ axis to  $(\eta-\eta_c)/\eta_c$ shows a data collapse. The noise at transition point $\eta_c(Pe)$ in this $Pe$ regime grows as an approximate power law $Pe^{0.2}$ as shown in the inset.
}
\label{NIPT2}
\end{figure}
We identify three different regimes, depending on how the function $S(\eta)$ behaves at different $Pe$ (Fig.\ref{NIPT1}).  
For $Pe\leq Pe_{\rm min}(=2.0)$, the curves fall on top of each other implying that a minimum activity $Pe_{\rm min}$ is  required for the system to show effects of $Pe$ in the macroscopic scale. 
This $Pe_{\rm min}$ vanishes with increase in system size(data not shown).
Increase in $Pe$, allows a single particle to cover larger space, nematically aligning with larger number of particles over a span of time. 
Thus a larger orientational noise $\eta$ is required to destabilize the nematic phase, increasing the transition noise $\eta_c$ with $Pe$. 
However, for a given density, there is a limit up to which $Pe$ can suppress the nematic to isotropic transition. This limit is decided by the available free space, determined by the mean density.  
For $Pe > Pe_{\rm max}(=20)$, again, the $S(\eta)$ curves fall on top of each other, making $\eta_c$ independent of $Pe$.
In between $Pe=Pe_{\rm min}$ and $Pe_{\rm max}$, the variations of $S$ with $\eta$ show data collapse when plotted against $(\eta-\eta_c)/\eta_c$~(Fig.\ref{NIPT2}). Within this restricted window of activity, we find a dependence of transition point $\eta_c$ with $Pe$ that shows an approximate scaling form $\eta_c(Pe) \sim Pe^{0.2}$~(inset in Fig.\ref{NIPT2}).

\begin{figure}[t]
\begin{center}
\includegraphics[width=8cm]{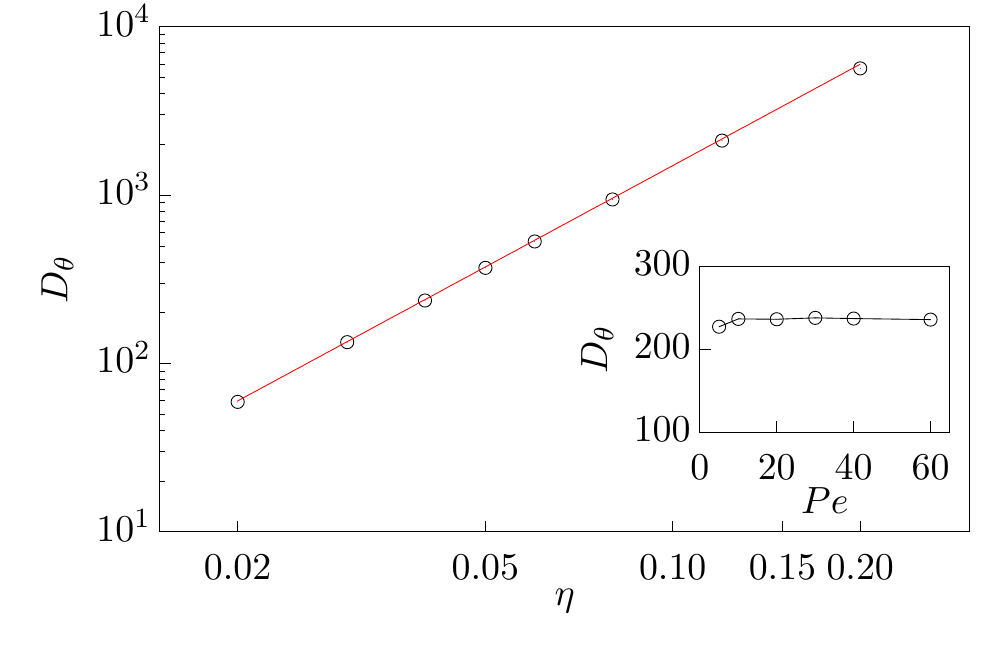}
\caption{(color online) The variation of the rotational diffusivity $D_{\theta}$ (black circles) as a function of the noise strentgh $\eta$ is shown for a fixed $Pe(=40)$. It shows a dependence of the form $D_{\theta} =  1.4\times 10^5\, \eta^2\, \t^{-1}$(red line). For a fixed $\eta$, $D_{\theta}$ doesn't depend on $Pe$ (shown in the inset for $\eta=0.04$).
}
\label{DR-Eta}
\end{center}
\end{figure}

\section{Orientational relaxation}
\label{sec_app:C}
In the presence of the nematic alignment, the active orientation of the particles undergo orientational relaxation controlled by the noise of strength $\eta$. In Fig.\ref{DR-Eta} we show that the orientational diffusion constant increases as $D_\h \sim \eta^2$, and remains essentially unchanged with variation of $Pe$.


\end {document}